\documentclass[journal]{IEEEtran}
\usepackage{cite}

\usepackage{url}

\usepackage[pdftex]{graphicx}
\graphicspath{{figures/}}
\ifCLASSOPTIONcompsoc
  \usepackage[caption=false,font=normalsize,labelfont=sf,textfont=sf]{subfig}
\else
  \usepackage[caption=false,font=footnotesize]{subfig}
\fi

\def \sys {\textsc{CrossCount}}

\usepackage{tabularx}
\usepackage{multirow}
\newcolumntype{Y}{>{\centering\arraybackslash}X}
\usepackage{tablefootnote}
\usepackage{xcolor,colortbl}
\usepackage{amsmath}

\ifCLASSOPTIONcompsoc
  \usepackage[caption=false,font=normalsize,labelfont=sf,textfont=sf]{subfig}
\else
  \usepackage[caption=false,font=footnotesize]{subfig}
\fi
\usepackage{hyperref}

\begin{document}
\title{\textsc{CrossCount}: A Deep Learning System for Device-free Human Counting using WiFi}
%
%
%

\author{
 Osama~T.~Ibrahim,~ Walid~Gomaa,~ and~Moustafa~Youssef
\thanks{Osama T. Ibrahim is with the Wireless Research Center, Egypt-Japan University of Science and Technology (E-JUST), Alexandria and the Computer Systems and Engineering Dept., Zagazig University, Zagazig, Egypt}
\thanks{Walid Gomaa is with the Computer Science and Engineering Dept., E-JUST and Alexandria University, Alexandria, Egypt}
\thanks{Moustafa Youssef is with Alexandria University, Alexandria, Egypt}
\thanks{Manuscript received April 19, 2005; revised August 26, 2015.}}

\maketitle

\begin{abstract}
 Counting humans is an essential part of many people-centric applications. In this paper, we propose \sys{}: an accurate deep-learning-based human count estimator that uses a single WiFi link to estimate the human count in an area of interest. The main idea is to depend on the temporal link-blockage pattern as a discriminant feature that is more robust to wireless channel noise than the signal strength, hence delivering a ubiquitous and accurate human counting system. As part of its design, \sys{} addresses a number of deep learning challenges such as class imbalance and training data augmentation for enhancing the model generalizability.

 Implementation and evaluation of \sys{} in multiple testbeds show that it can achieve a human counting accuracy to within a maximum of $ 2 $ persons $ 100\% $ of the time.  This highlights the promise of \sys{} as a ubiquitous crowd estimator with non-labour-intensive data collection from off-the-shelf devices.
\end{abstract}


%
\IEEEpeerreviewmaketitle

\section{Introduction}
 %
 %
 %
 
 \label{Introduction}
 
 \IEEEPARstart{M}{onitoring} the human count in a given area of interest is a crucial part in many pervasive applications such as smart guiding in museums, energy management in smart buildings, indoor analytics, and people evacuation in emergency situations. For example, in a retail store, lightening and air conditioning can be automatically adjusted based on the clients' density in each section. Furthermore, the occupancy statistics can assess the store sections that attract more visitors to plan for future business~\cite{retail}.

 Due to its importance, human-counting has attracted the attention of the research community. For example, computer vision researchers, aided with the recent advancement in deep learning, presented high-accuracy human counting systems~\cite{CrowdNet,CNN,CrossScene,RCNN}. Images/videos captured by cameras from the area of interest are processed to estimate the human count. Most of these systems are built using Convolutional Neural Networks (CNN) with various architectures and feature optimization techniques. After the network is trained with a large set of labeled images/videos, the human density is estimated by processing images/videos of the crowd using the trained net. However, vision-based systems require high installation cost, suffer from blind spots and occlusion issues, require high computational power, are limited in functionality in poor lightening conditions, and raise privacy concerns. In addition, they cannot work through the walls.  This \emph{through-wall} capability is highly-desirable in a number of applications such as law enforcement~\cite{hunt2001}.

 To address these issues, algorithms based on analyzing  the RF signals have been introduced. In particular, they analyze the signal received from the already-installed wireless infrastructure.  RF-based systems can be either device-based or device-free. In device-based systems, each human target must be equipped with a device, such as a cell phone~\cite{WiCounter,deviceBased}, which limits the system ubiquity. On the other hand, in device-free systems, the number of human targets inside an area is estimated by analyzing their impact on the wireless links covering this area, without requiring them to carry any device. The concept of RF device-free localization was introduced in 2007~\cite{challenges}, presenting human counting as one of the challenges that faces the newborn technique. Since then, a number of device-free counting techniques have been introduced based on different features and machine learning algorithms~\cite{nakatsuka2008,Nuzzer,yoshida2015,Depatla2015}.
 
 In an early study, Nakatsuka et al.~\cite{nakatsuka2008} demonstrated the feasibility of using Received Signal Strength (RSS) of radio links to estimate human counting. The authors showed empirically that increasing the number of people leads to higher variance in the RSS signal of a single RF link. Based on that, they derived a linear formula that relates the human count to the RSS average and variance. The counting functionality in the Nuzzer system~\cite{Nuzzer} extended this model to work on a large scale. Based on observations, the authors showed that the variance of a single link is not enough to differentiate clearly between the human count classes. Therefore, they proposed to use the average relative variance of $12$ WiFi links to count only up to two persons with $81\%$ accuracy and up to three persons with $75\%$ accuracy. 
 Adding more complication to the model, Yoshida et al.~\cite{yoshida2015} examined non-linear regression to capture the relation between the RSS and the number of people. Specifically, they used a Gaussian kernel to perform regression of RSS absolute values of 10 wireless links. Depatla et al.~\cite{Depatla2015} propose a probabilistic approach to calculate the RSS probability mass function (PMF) of one link as a profile for each case of human count. When testing using an RSS vector, they compare its PMF with the pre-calculated profiles and report the nearest one, achieving $25\%$ exact counting accuracy with estimation error of $ 2 $ or less $ 63\% $ of time.
 
 The main limitation of the above approaches is that the area of interest should be covered by a large number of WiFi links in order to achieve an acceptable counting accuracy, which is not the case in many wireless environments and applications. One possible solution to resolve this trade-off between links density and counting accuracy is to use channel state information (CSI) instead of RSS~\cite{ElectronicFrogEye,TrainedOnce,Cheng2017}, where the data of all RF sub-carriers of every WiFi link is available for processing. Unfortunately, unlike RSS, reading CSI data is not widely supported by all wireless cards. In addition, both RSS- and CSI-based techniques cannot work-well in through-the-wall scenarios, due to the attenuation of the RF signal.
 
 Recently, Depatla et al.~\cite{Depatla2018} proposed to utilize the WiFi link blockage events as a discriminate feature instead of features depending on the RSS exact values. In particular, the system embeds the inter-arrival times between the link line of sight (LoS) blockages into a renewable stochastic process that models the human motion mathematically. In addition to providing through-wall counting, the blockage pattern performs well in case of counting moving targets; which incur more RSS variance affecting the estimation model. However, this system accuracy significantly degrades in a number of real-world scenarios, as we quantify in Section~\ref{secEvaluation}, due to the oversimplified assumptions in the used mathematical model. These simplifications include discarding the order of inter-arrival times which is an important part of the context information and simulating the human motion to generate the model training data. Besides, the proposed mathematical model is tailored for special cases of testbeds where the WiFi link is in the middle and aligned with the area of interest. For any different setup, the mathematical model in~\cite{Depatla2018} does not fit leading to deteriorated performance.
 
 In this paper, we present \sys{}, a through-wall human counting system that leverages a Recurrent Neural Network (RNN) to map a sequence of link inter-blockage temporal pattern to the human count using a single WiFi link. The idea is that, the higher the number of people in an area of interest the shorter the time between blocking a single WiFi link and vice versa. As part of \sys{} design, we introduce different modules to address practical issues such as reducing the labour-intensive calibration required for training a deep learning model as well as handling the imbalance in the number of training cases between the different counting classes.
 
 Implementation and evaluation of \sys{} in different testbeds show that it can provide the exact human count $59\%$ of the time. This increases to $ 100\% $ to within two persons difference in count. This is achieved  using the information of only a single WiFi link, highlighting \sys{} promise as a ubiquitous through-the-wall human counting  system. To sum up, the main contributions of this work are threefold:
 \begin{itemize}
    \item We present the architecture and details of \sys{}: a deep learning system that leverages the temporal blockage information of a single WiFi link to provide accurate device-free human counting.
    \item Beneath the folds of \sys{}, we propose a novel technique for training data augmentation and class balancing to significantly decrease the data collection overhead.
    \item We implement the \sys{} and thoroughly evaluate its performance in clear and cluttered WiFi testbeds by counting up to $ 10 $ persons.
 \end{itemize}
 
 The rest of the paper is organized as follows.  Section~\ref{secOverview} provides an overview on how \sys{} works. Section~\ref{secDetails} gives the details of \sys{} components and how it deals with different practical challenges. We evaluate the system performance in Section~\ref{secEvaluation}. Finally, Section~\ref{secConclusion} concludes the paper and discusses future directions.
 

\section{System Overview}\label{secOverview}
 \begin{figure*}
     \centering
     \includegraphics[width=\textwidth]{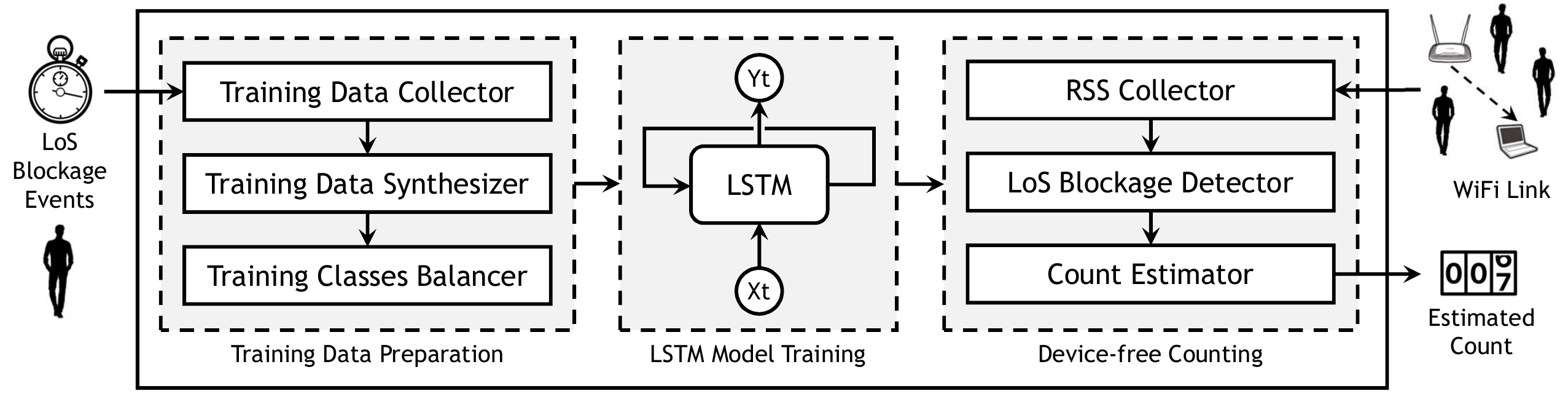}
     \caption{\sys{} system architecture.}
     \label{figSystemArchitecture}
 \end{figure*}
 
 In this section, we present a typical scenario of how \sys{} works and give an overview about the information flow through its main modules. The details of these modules are described in subsequent sections. We assume an indoor area covered by a single WiFi link whose transmitter and receiver are behind the walls. There is an unknown number of people that are moving casually inside this room. Given only the RSS readings over time at the receiver, our goal is to infer the human count inside this area of interest.
 
 The \sys{} system architecture is depicted in Fig.~\ref{figSystemArchitecture}. The basic idea of \sys{} is to map a sequence of link-blockage time to the estimated count based on the observation that the more people in the area of interest the shorter the time between link blockages. To do this mapping, \sys{} leverages a recurrent neural network. In particular, for discretized time, the input sequence to the RNN is a stream of binary values which are $1$ at the time instances when an LoS blockage event encountered and $0$ otherwise. The output of the RNN is the estimated count.
 
 \sys{} operates in three stages: preparing the training data, training the deep network, and finally classifying the input sequence by leveraging the trained network.  During the first stage, the training sequences are collected \textit{\textbf{manually}} in a light-weight process by recording the timestamps of the WiFi link virtual LoS blockages by a \textit{\textbf{single}} walking human \textit{\textbf{without reading the link RSS}}. However, during the online counting stage, the link-blockage events are calculated \textit{\textbf{automatically}} by processing the captured WiFi link RSS stream at the receiver.
 
 The typical way of collecting the training data in literature is to try all the combinations of the number of people in the area of interest~\cite{nakatsuka2008,yoshida2015,ElectronicFrogEye,TrainedOnce,Cheng2017}. This makes the training task labor intensive and limits the systems ability to detect a large number of people~\cite{RadioGrapher}. To address the scale and overhead of collecting the training data, \sys{} introduces a new technique that depends on collecting the training samples using \textbf{only a single person}. This single-person training data is then processed to automatically generate the training sets for all the other count classes. Specifically, during the \textit{Training Data Preparation} stage, the \textbf{Training Data Collector} module records the LoS blockage events caused by a \textit{single} person moving in the environment to generate the training data for the single-human count class. Thereafter, the \textbf{Training Data Synthesizer} module processes the collected data to generate the training set for multiple-user classes.

 The generated data also suffers from the imbalanced data distribution for the different count classes.
 In addition, deep learning models require large amounts of training data. The \textbf{Training Classes Balancer} module tackles both of these issues by augmenting the training set with synthesized data. This also enhances the system generalizability and increases its ability to deal with the noisy characteristics of wireless channels.

 \sys{} then trains a Long Short-Term Memory (LSTM) RNN network  to characterize the blockage pattern of each human count class. The data preparation and training phases are done offline only once while deploying the system.
 
 During the online system operation in the \textit{Device-free Counting} stage, \sys{} estimates the unknown crowd count by extracting their LoS blockages sequence pattern and classifying it using the trained LSTM model. Specifically, the \textbf{RSS Collector} reads the signal strength received over the WiFi link for a specific time window. The \textbf{LoS Blockage Detector} module processes the RSS values and estimates the LoS blockage events timing encountered along this window in quantized time units. This generates a binary link blockage sequence that has a 1 in the time slot that the link was blocked inside the current window and zero otherwise. The input binary sequence is then passed to the \textbf{Count Estimator} module which uses the learned LSTM network to estimate the human count.
 
\section{The \sys{} System}\label{secDetails}
 In this section, we provide the details of the main \sys{} modules, including diminishing the training data collection overhead, class balancing, and enhancing  the system generalizability. We start by an overview of the \sys{} training process.
 
 \subsection{\sys{} Training Process Overview}
   As a supervised machine learning system, each sample in the \sys{} training set is sequence of the link blockage times which is labeled with the number of persons generating this sequence. The link blockage time is mapped to a bit map, where ones represent instances of time the link was blocked in a discrete time slotted system (Fig.~\ref{figSuperposition}).
   
   The direct way to collect these samples is for the specified number of persons to move inside the area of interest for a certain time window, $w$, record the time instances they cross the LoS, and label the generated sequence with the human count. This should be repeated for each and every crowd count class, i.e. number of humans inside the area of interest. Moreover, each class should have a large number of training samples to generate a well-trained deep network.
 
   Accordingly, collecting the training data is a labor-intensive and time consuming task. Some previous work, e.g.~\cite{Depatla2015,Depatla2018}, proposed a human motion model to simulate this task. However, the mathematical model assumes simple motion that does not capture real life complex scenarios, affecting the system accuracy as we quantify in Section~\ref{secCompareSystems}. \sys{} resolves this challenge by collecting the training samples for only a \textit{single} user, form which the whole training data for any arbitrary number of humans inside the area of interest is extrapolated.
 
   \subsubsection{Training Data for Single Target}
    The \textit{Training Data Collector} module allows a single person to move randomly inside the area of interest while recording the timestamp each time she crosses the WiFi link virtual LoS. Note that the training process does not require processing the RSS of the link as it is based on visually recording when the user crosses the virtual line between the transmitter and receiver. This is because \sys{} depends only on the time of link blockage and not on the specific RSS.
    
    Assuming a discrete time space, a training sample takes $ w $ time steps, to be collected.   This is repeated $m$ times to collect different training sequences, each of length $ w $. We call these manually collected samples the ``\textit{the original training samples}''.

   \subsubsection{Reducing Training Data Collection Overhead for Multiple Persons}
    \begin{figure}
     \centering
     \includegraphics[width=\columnwidth]{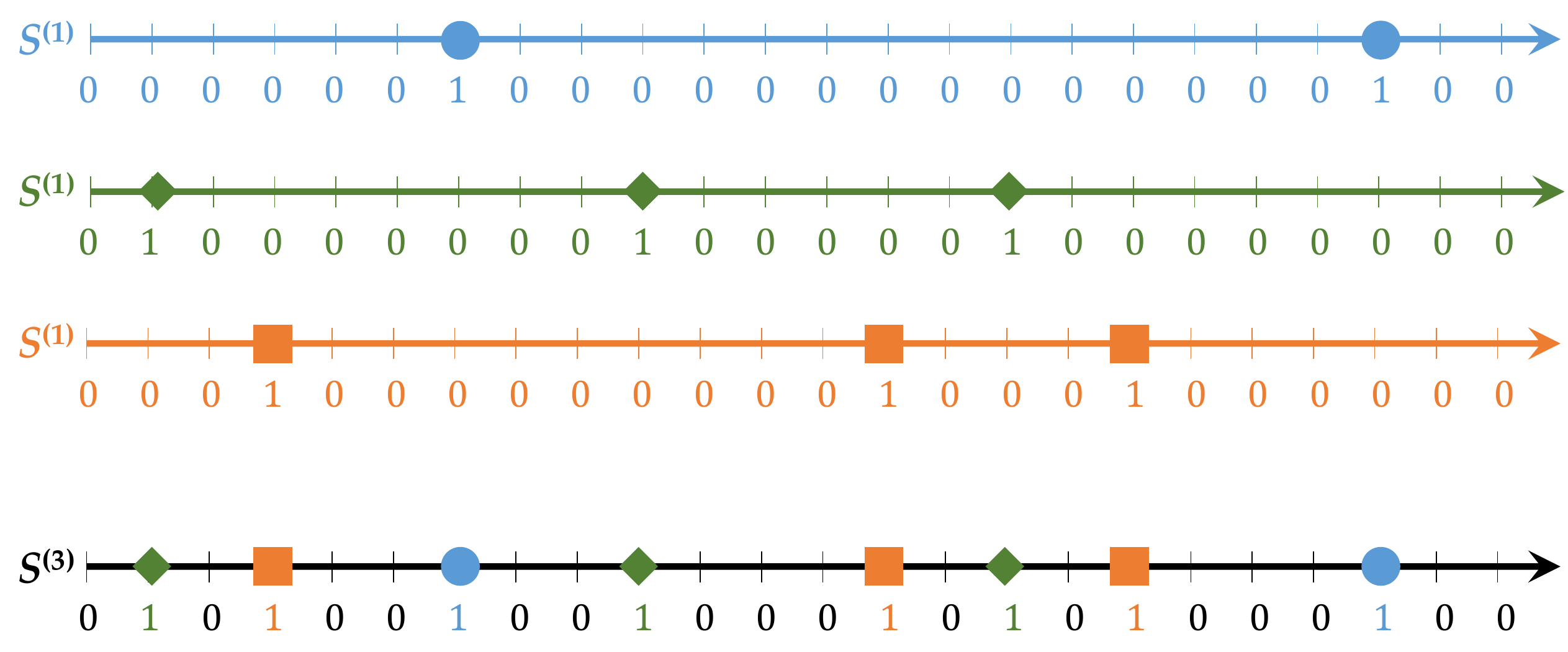}
     \caption{An example describing the superposition of $ 3 $ single-person sequences $( S^{(1)} )$ to generate a $ 3 $-persons sequence $( S^{(3)} )$. The circles, diamonds, and squares represent the LoS blockages of the first, second, and third sequence respectively.}
     \label{figSuperposition}
    \end{figure}
  
    \sys{} generates the training samples for the \textbf{multi-person} classes using the collected data from the \textbf{single-person} case as described in the previous section. To do that, we assume that each person is moving independently of the others. As a result, the LoS blockage sequence of a user is independent from the others and the blockage pattern for \textbf{multiple} persons can be calculated as the \emph{superposition} of the individual persons' blockages. In particular, \sys{} randomly selects $ k $ sequences out of the $ m $ collected samples of a single person and consider each of them as a link blockage sequence for  a different user. The $ k $-person count class training sample is synthesized from the superposition of these $ k $ sequences. Fig.~\ref{figSuperposition} shows an example on how to generate a training sequence for $ 3 $ persons from $ 3 $ single-person training sequences. For each time instance, the superposed sequence reports an LoS blockage if any of the single-person sequences encounters a blockage at this instance. For each multi-person count class, the \textit{Training Data Synthesizer} module applies the superposition technique over all the available combinations of the $ m $ original training samples to generate the higher count classes training set.

   \subsubsection{Handling the Class Imbalance Problem\label{secDataSynthesis}}
  
   \begin{table}
    \caption{An example on the imbalanced training data for 0-5 count classes. Note that the zero-count class has just one example represented by the bit sequence of all zeros, i.e no link blockage events in an empty area of interest.}
    \label{tabImbalanced}
    \begin{tabularx}{\columnwidth}{| l | X | X | X | X | X | X |}
     \hline
     \textbf{Count Class} & $ 0 $ & $ 1 $ & $ 2 $ & $ 3 $ & $ 4 $ & $ 5 $\\
     \hline
     \textbf{No. of Samples} & $ 1 $ & $ 15 $ & $ 105 $ & $ 455 $ & $ 1365 $ & $ 3003 $\\
     \hline
    \end{tabularx}
   \end{table}
  
    The \textit{Training Data Synthesizer} uses the superposition technique to generate $ \binom m n $ training samples for each $ n $-count class. For instance, Table~\ref{tabImbalanced} lists the size of the training set for the $ 0 $ to $ 5 $-count classes as generated from $ m = 15 $ original training samples. There is only one training sample for the $ 0 $-count class which is a sequence of zeros as it does not encounter any LoS blockage during the counting window. The table shows that the class distribution is not uniform and is severely skewed. This imbalanced class data problem is well-known in machine learning~\cite{Imbalanced, hamada} and leads to a bias in the classifier towards the classes with large training data.
    
    Traditionally, random oversampling and sub-sampling are simple and well-established solutions for this class imbalance problem~\cite{Imbalanced}. Oversampling increases the minority classes by replicating the training samples, while sub-sampling decreases the majority classes by randomly removing some of its samples. Oversampling leads to overfitting the training data~\cite{Overfitting}. Therefore, in \sys{}, the \textit{Training Data Balancer} module sub-samples the majority classes training set. In addition, \sys{} also augments the data from the minority classes by injecting noise into the training samples in order to improve the algorithm generalization capability and system robustness~\cite{BookMIT}. Noise is injected by randomly flipping some of the bits in the link-blockage bit steam input. This simulates injecting false link-blockage events or missing an actual link-blockage event, which may occur in reality due to the noisy wireless channel. This noisy training data increases \sys{} robustness and avoids overfitting.
    
    The Training Data Synthesizer and Class Balancer modules are implemented as follows. The main input from the Training Data Collector is the $ m $ manually collected original single-person sequences, from which the multiple-persons training data is synthesized. Each training sequence is $ w $ seconds in length. For the zero-entity class, there is only one original training sample which is a sequence of zeros with length $ w $, expressing that no LoS blockage happened along the time window. This sequence is noised by flipping any single bit along the sequence length. This generates a total of $ w+1 $ training sample for the $ 0 $-count case, limiting all other classes to this size of training samples, to ensure a balanced training set. The training data for any other $ n $-count class is generated as follows. A random combination of $ n $ original sequences is selected from the $ m $ inputs and a superposed training sample is synthesized from the bitwise logical OR of all the randomly-selected sequences. Another superposed sample can be generated by selecting another random sequence combination. We keep repeating this till meeting the class balancing condition where the number of generated sequences is equal to $ w+1 $. If all the available $ \binom{m}{n} $ sequence combinations are processed before meeting the balancing condition, the training set for the current $ n $-class is augmented by data noising. A new training sequence is generated by selecting a random superposed sequence, flipping any random bit inside, and appending it back to the training set.
    
   \subsection{Model Training}
   \begin{figure}
    \centering
    \includegraphics[width=\columnwidth]{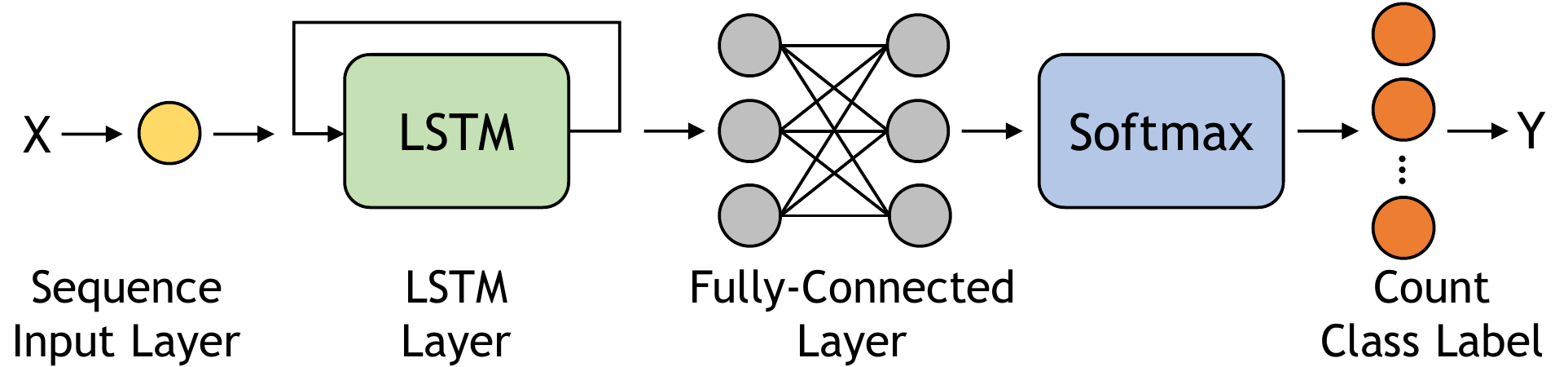}
    \caption{\sys{} network diagram.} 
    \label{figLSTM}
   \end{figure}
   
    Recurrent Neural Networks (RNN) model the contextual information in the input sequences. RNNs have been employed in many applications whose input data are sequences of features such as speech recognition~\cite{speechRec} and handwriting alignment~\cite{handWriting}. However, when dealing with long sequences, traditional RNNs suffer from the vanishing gradient problem during training~\cite{BookPhD}. Long Short-Term Memory (LSTM)~\cite{LSTM} architecture is one of the most commonly-used solutions to this problem. Therefore,  \sys{} leverages them to capture the contextual information in the link-blockage binary input sequence. Figure~\ref{figLSTM} shows the \sys{} LSTM architecture. The input sequence is one temporal bit stream generated from the deployed WiFi link, representing the link-blockage events as described in Section~\ref{secDetails}. Therefore, we employ a one-dimensional input layer. Each crowd count is an output class of the network. Accordingly, the size of output layer is determined and preceded by a fully connected softmax for classifying among more than two output classes. The LSTM layer implements the tanh and sigmoid functions as the default configuration for cell state and gates activation respectively. The output layer implements cross entropy loss function.
    
    During the Model Training phase, the LSTM network is trained using the original and synthesized training samples. The synthesized data allows the model to generalize better to noisy unseen data.
    
   \subsection{Online Phase: Human Counting}
    In this stage, the trained network is used to estimate the unknown crowd count. The WiFi link RSS stream is analyzed to extract the LoS blockage pattern of the current crowd. Feeding this blockage sequence to the trained LSTM activates a forward path to an output class which will be reported as the count estimate.
    
   \subsubsection{The RSS collector}
    The RSS reader is a lightweight agent running on the receiver to record the temporal change of the Received Signal Strength (RSS) from the transmitter along with its time stamp. Finally, a sequence of $ l $ timestamped readings, $ \{ (r_0,t_0), (r_1,t_1), ..., (r_l,t_l) \} $, is collected and sent to preceding modules. Where each reading is streamed as a pair of the RSS value and time $ (r,t) $.
  
   \subsubsection{LoS Blocking Detection}\label{secBlockingDetection} 
    \begin{figure}
    \centering
    \subfloat[Single Person.\label{figStream1Person}]{
     \includegraphics[width=0.97\columnwidth]{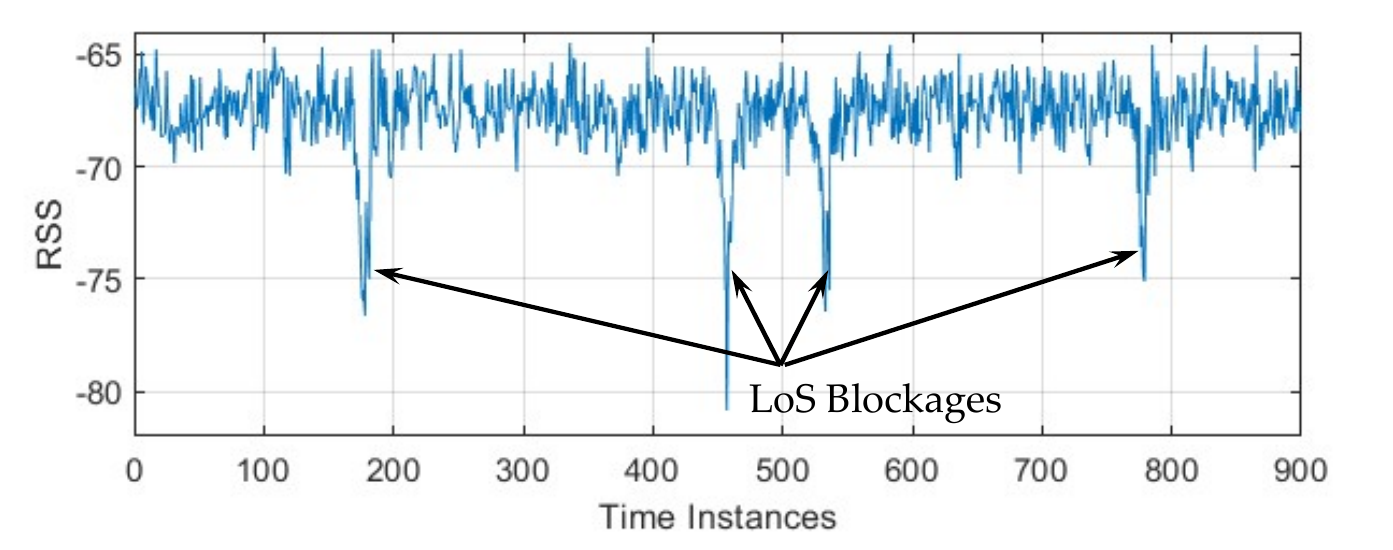}
    }
     
    \subfloat[Three Persons. For visual clarity, not all the blockages are indicated.\label{figStream3Person}]{
     \includegraphics[width=0.97\columnwidth]{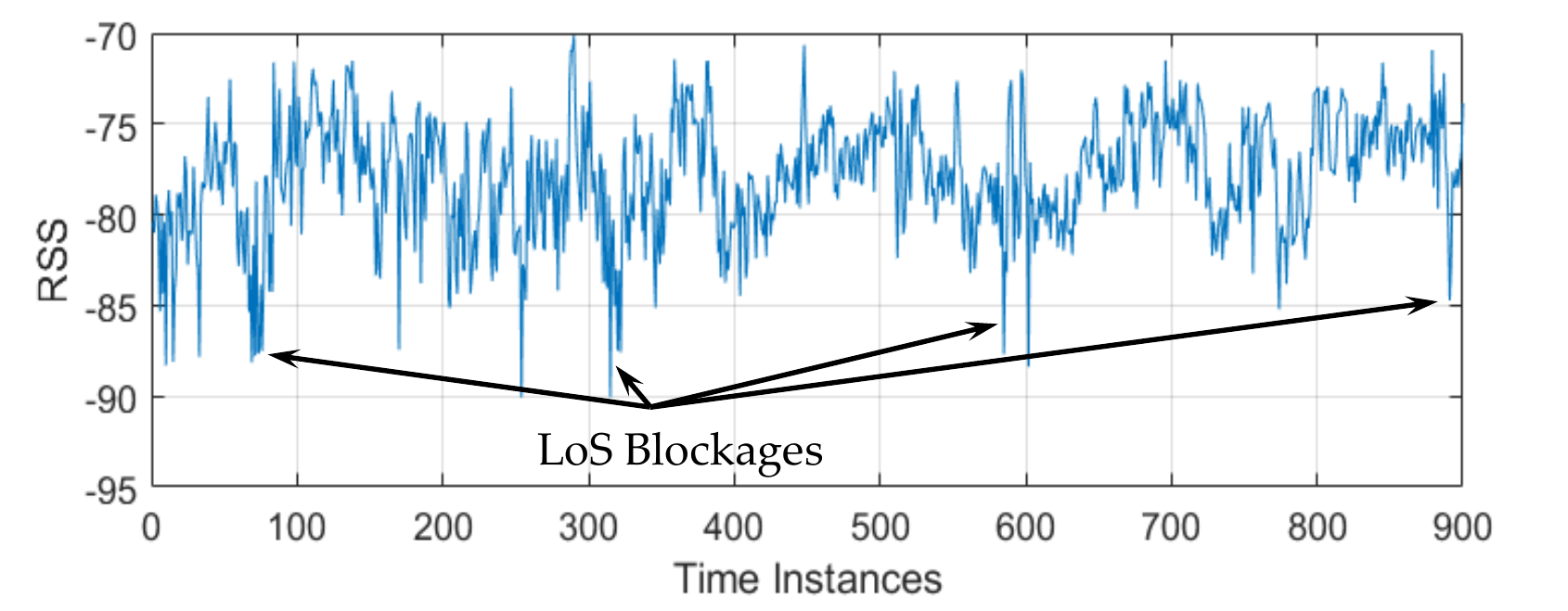}
    }
    \caption{The RSS measurements of a WiFi link while some persons are moving around. The time instances when a person crosses the LoS are highlighted.}
     \label{figStream}
    \end{figure}
    
    Fig.~\ref{figStream} shows a typical example of the RSS measurements of a deployed WiFi link while a human is moving around. The figure shows that the RSS measurements are significantly attenuated when someone crosses the LoS. The figure shows that there is a down pulse whenever the target person crosses the link LoS while the other fluctuations are due to the multipath effect~\cite{Depatla2018}. From the above observations, the RSS fluctuations due to multipath are limited within a certain level around the mean value, while the LoS blocking has a higher down pulse. Accordingly, the \textit{LoS Blockage Detector} of \sys{} captures an LoS blockage by a simple thresholding process on the RSS changes. Specifically, at any time instance if the RSS exceeds its mean value $ \overline{r} $, by a certain threshold, $ \tau $, an LoS blocking event is declared at this instance. 
    Finally, the module converts the detected LoS blockage events into a binary stream  $ B=\{b_0, b_1, ..., b_{w-1} \} $, where 
      \begin{equation}
        b_i = \left\{ \begin{array}{lc}
        1 & \mbox{if} \quad  r_j \geq \overline{r} + \tau, \, T_{i} \leq t_j < T_{i+1} \\
        0 & \mbox{otherwise}
        \end{array}\right.,
      \end{equation}
      \begin{equation}
        \overline{r} = \frac{1}{l}\sum_{k=0}^{l-1}r_k 
      \end{equation} Where $ T_i $ is the $ i $th time step.
      
   \subsubsection{Count Estimation}
    The crowd count in the current counting window is estimated by classifying its blockage pattern. The sequence $ B $ is fed forward to the trained LSTM, and eventually, \sys{} reports the output class as the human count estimate.
 
  \section{Evaluation}\label{secEvaluation}
    In this section, we evaluate the \sys{} performance. We start by describing the experimental testbeds and training process, followed by testing the effect of system parameters on counting accuracy. Finally, we compare \sys{} performance to the state-of-the-art RF human counting systems. Table~\ref{tableDefSysParam} contains the default system parameters values used throughout the evaluation section.
 
 \begin{table}
   \caption{Default system parameters.}
     \begin{tabularx}{\columnwidth}{|X|c|c|}
      \hline
      \textbf{Parameter} & \textbf{Range} &
      \begin{tabular}{@{}c@{}}\textbf{Default value} \\ \textbf{(Testbed $1,2$)}\end{tabular} \\
      \hline\hline
      \textbf{Blockage Detection Threshold ($ \tau $) (dBm)} & $ 0 $ - $ 10 $ &  $ 5 $, $ 5.5 $\\
      \hline
      \textbf{Estimation Window ($ w $) (min.)} & $ 1 $ - $ 5 $ & $ 5 $ \\
      \hline
      \textbf{LSTM layer size (units)} & $ 10 $ - $ 100 $ & $ 100 $\\
      \hline
      \textbf{Number of training epochs} & $ 10-150 $ & $ 120, 150 $ \\
      \hline
      \textbf{Training mini-batch size (samples)} & $ 1-30 $ & $ 15, 3 $ \\
      \hline
     \end{tabularx}
     \label{tableDefSysParam}
    \end{table}
 
   \subsection{Experimental Testbeds}
   \begin{figure}
   \centering
    \subfloat[Testbed 1.\label{figTestbed1}]{
     \includegraphics[width=0.6\columnwidth]{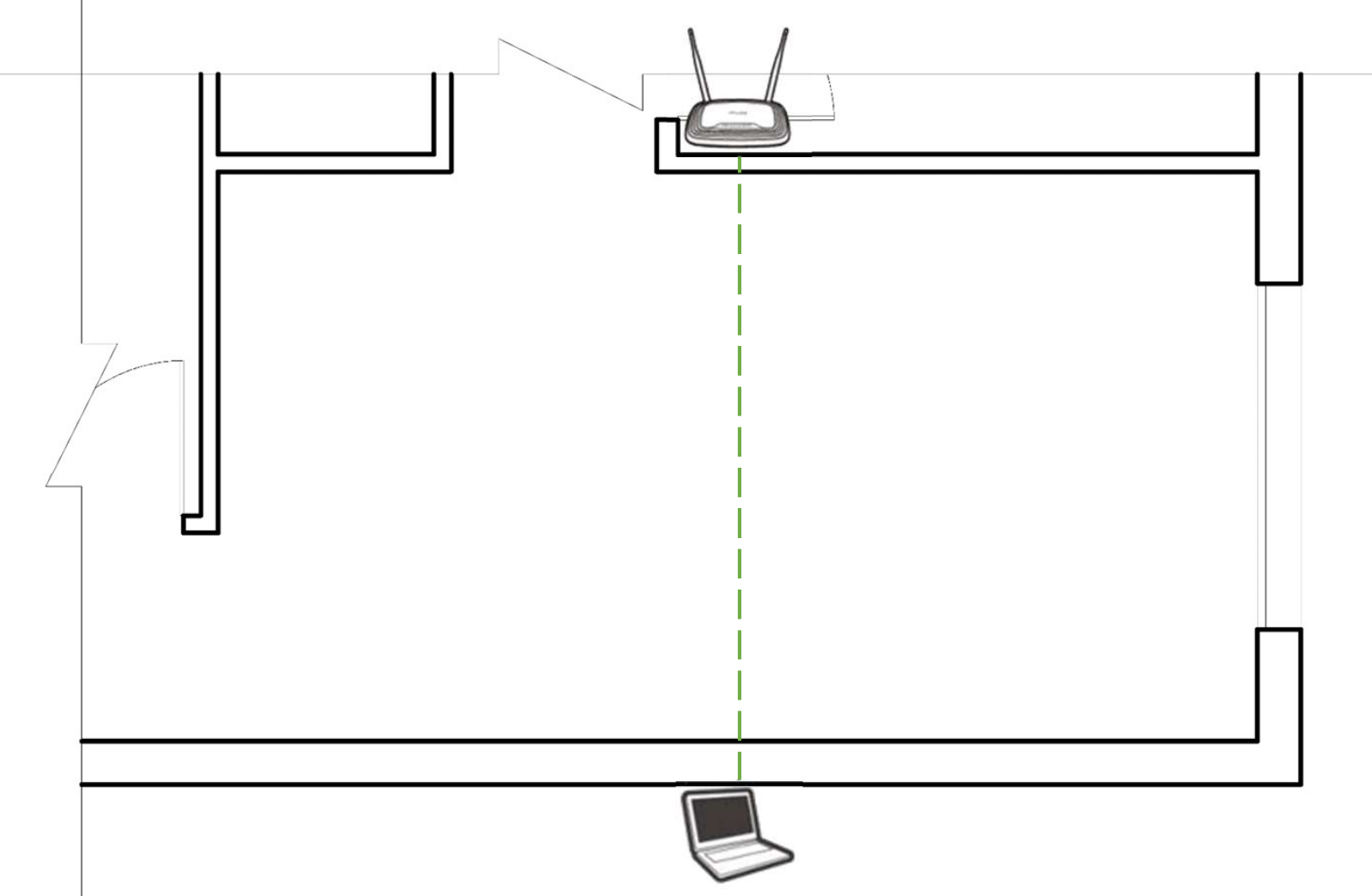}
     }
     
    \subfloat[Testbed 2\label{figTestbed2}]{%
     \includegraphics[width=0.85\columnwidth]{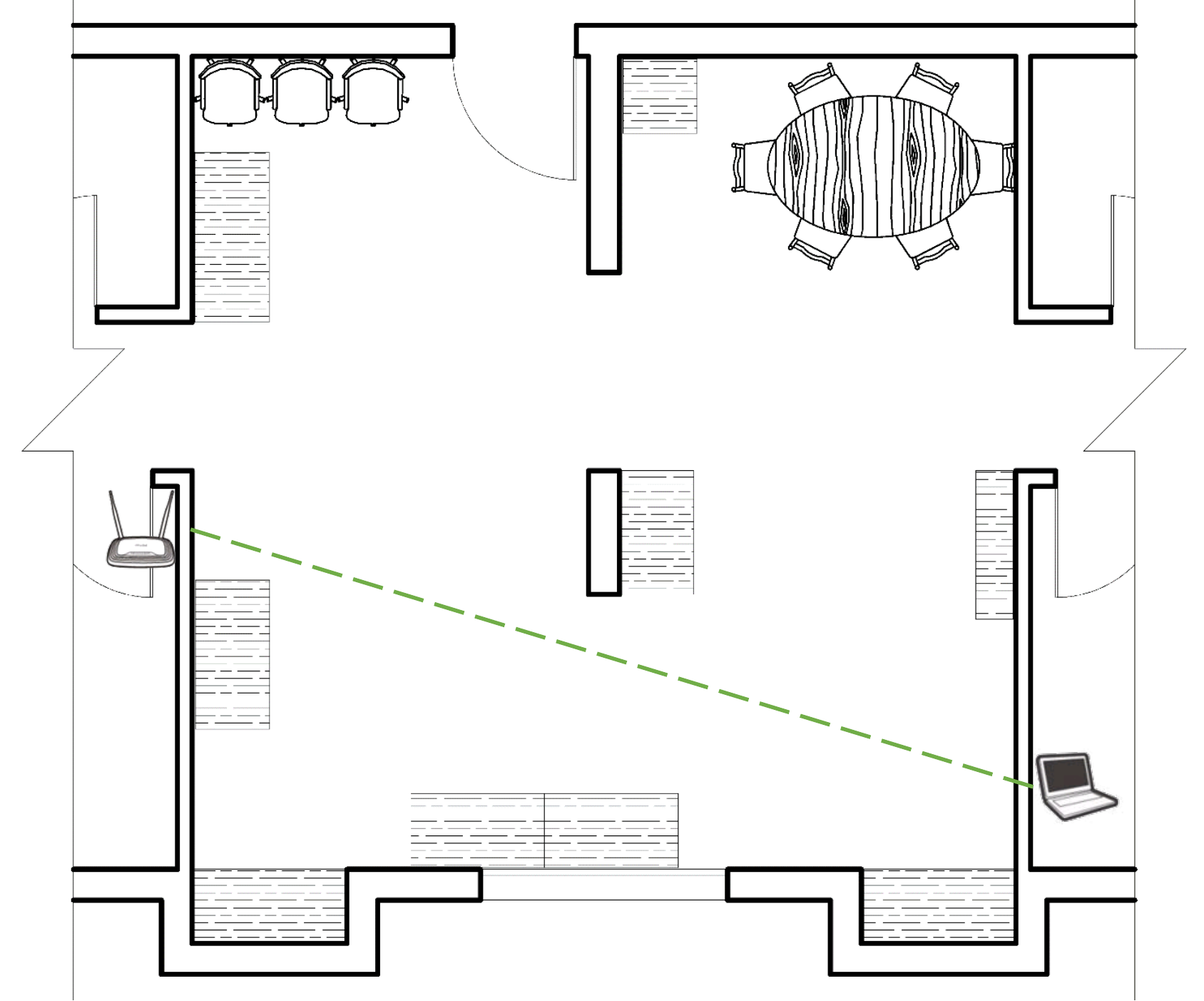}
     }
     \caption{Evaluation Testbeds.}
  \end{figure}

    We evaluated \sys{} in two testbeds. The first one is a $ 19.8 $ m$ ^2 $ room in our lab as shown in Fig.~\ref{figTestbed1}; it is a controlled environment with low multipath where the WiFi link is deployed right in the middle of the room. This is similar to the testbed used in \cite{Depatla2018}. The second testbed is more complex, as shown in Fig.~\ref{figTestbed2}: it is a larger hall of $ 50 $ m$ ^2 $, rich in multipath due to its furniture, and the WiFi link is not aligned with the hall nor in the middle. In both testbeds, the transmitter is mounted to the wall at height of $ 185 $ cm, the receiver is placed $ 75 $ cm from the floor, and the walls are made of bricks.
 
   \subsection{Training Data}
    The original training set is collected using a \textit{single} person who walks in each testbed for $ 100 $ minutes generating a $ 100 $-minutes-long sequence. This sequence is divided into sub-sequences of $ w $-length each. Following the data synthesis process described in Section~\ref{secDataSynthesis} with the default counting window, $ 301 $ training sequences are generated per count class resulting in a total training set of $ 3.3k $ samples when testing using $ 10 $ persons.  The details and types of the generated training sequences are listed in Table~\ref{tabTrainData}.
    
    The \sys{} model was trained using stochastic gradient descent with momentum (SGDM) optimizer, with $ 0.9 $ momentum and $ 0.01 $  initial learning rate. On average, the model training process takes us $ 5 $ hours to finish on a Dell XPS 8500 desktop with a core i7 processor and $ 12 $ GB memory.

  \begin{table}
   \caption{Collected Raw data for Training Sequences}
   \label{tabTrainData}
   \begin{tabularx}{\columnwidth}{| l | X | X | X | X |}
   \hline
   \textbf{Count Class} & $ \textbf{0} $ & $ \textbf{1} $ & $ \textbf{2} $ & $ \boldsymbol{\geq} \textbf{3} $ \\
   \hline\hline
   \textbf{Collected Sequences} & $ 1 $ & $ 20 $ & $ 0 $ & $ 0 $\\
   \hline
   \textbf{Synthesized Sequences} & $ 0 $ & $ 0 $ & $ 190 $ & $ 301 $\\
   \hline
   \textbf{Generated Noisy Sequences} & $ 300 $ & $ 281 $ & $ 111 $ & $ 0 $\\
   \hline
  \end{tabularx}
  \end{table}

  \subsection{Effect of System Parameters}
   In this section, we evaluate the effect of changing the system parameters on the counting performance as reported by the Absolute Counting error $ \varepsilon $, where $ \varepsilon $ is the summation of the absolute difference between real and estimated counts in all cases.
   \begin{equation}
     \varepsilon = \sum \left | C_r - C_e \right |
   \end{equation}
   Where $ C_r, C_e $ are the real and estimated counts respectively.
   
   \subsubsection{LoS Blockage Detection Threshold}
     
    \begin{figure}
    \centering
    \includegraphics[width=0.9\columnwidth]{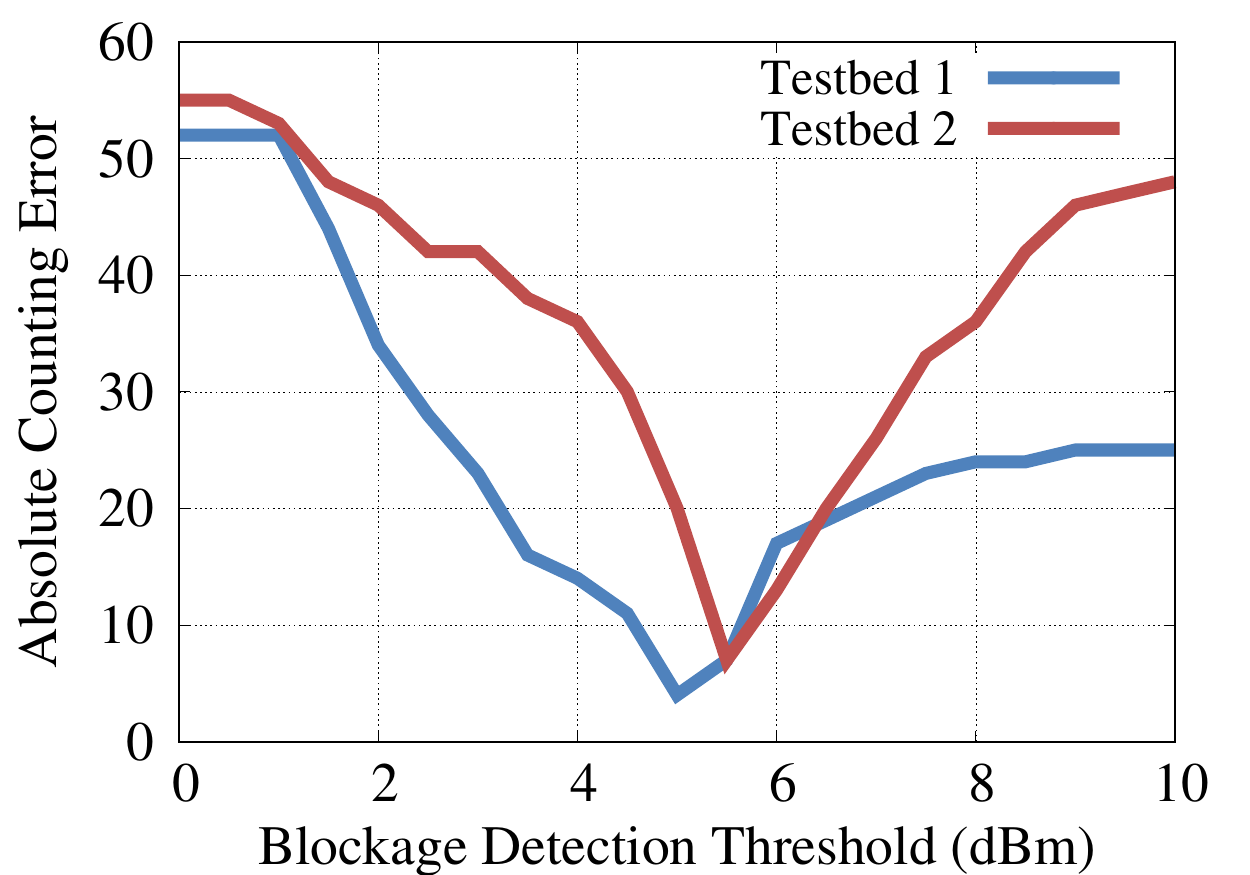}
    \caption{Evaluation of blockage detection threshold. }
    \label{figEvaluThreshold}
    \end{figure}
   
    Fig.~\ref{figEvaluThreshold} shows the absolute counting error at different  values of the LoS blockage detection threshold ($ \tau $) in both testbeds. For small thresholds, any minor fluctuation in RSS values is falsely detected as an LoS blockage; this increases the LoS blockage rate and reports extra persons in the testbeds. In contrast, high thresholds lead to missing real LoS blockage events,  underestimating the actual crowd number and increasing the absolute error. Throughout the rest of the evaluation section, we set $ \tau = 5 $ and $ 5.5 $ as the default blockage detection thresholds for testbeds $ 1 $ and $ 2 $ respectively as they lead to the best performance.
  
   \subsubsection{Counting window length}
 
   \begin{figure}
    \centering
    \includegraphics[width=0.9\columnwidth]{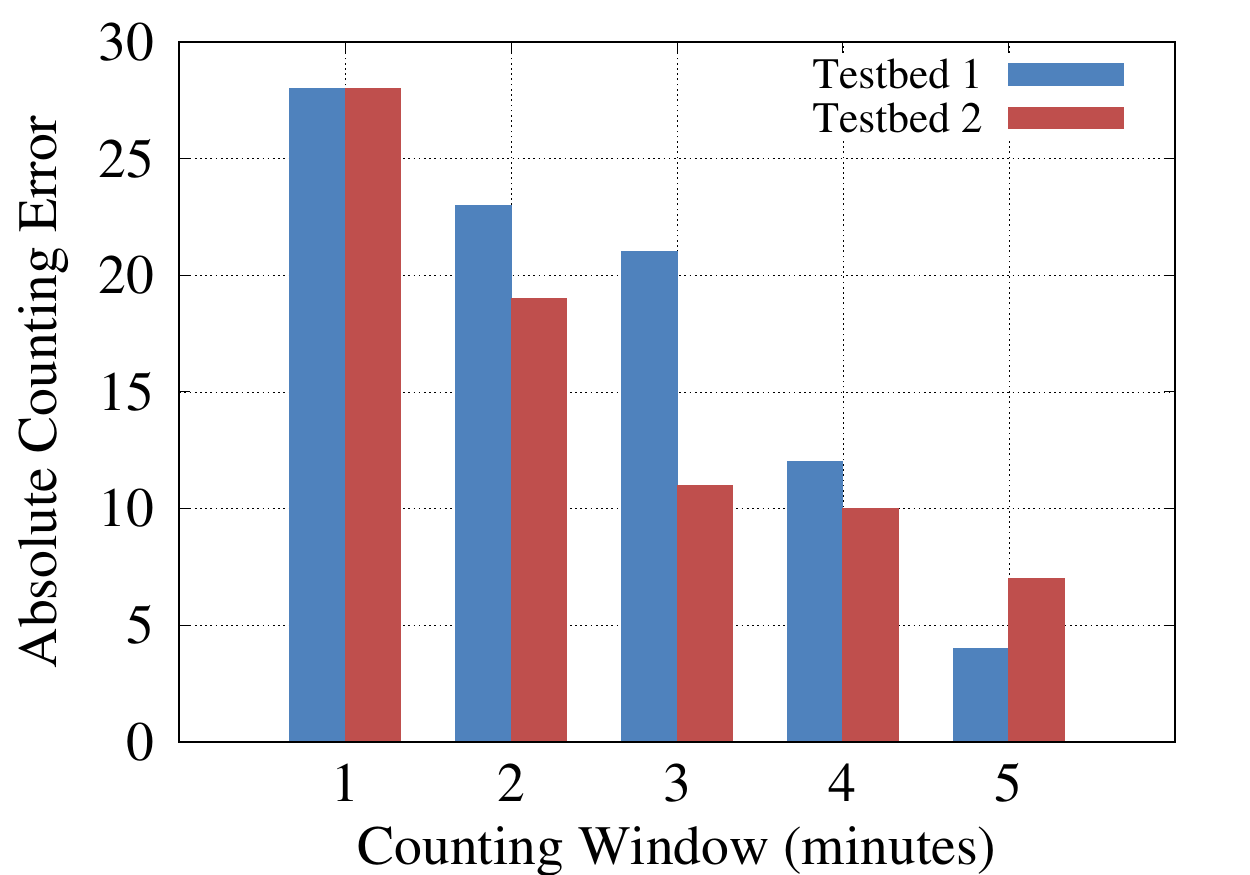}
    \caption{Evaluation of the counting window length.}
    \label{figEvaluWindow}
   \end{figure}
  
    Figure~\ref{figEvaluWindow} shows the effect of various time windows $ w $ on the counting performance. For short windows, the information reflected by the blockages sequence encountered within the window is not enough for providing accurate count estimates. Increasing the window size to $ 5 $ minutes, as used in this paper, gives the best counting accuracy. Note that this is the same estimation delay used in  literature~\cite{Depatla2015,Depatla2018,yoshida2015}. The system designer needs to tune this parameter to trade-off latency and accuracy of estimation based on her specific application need.
    
   \subsection{Comparisons with other systems}\label{secCompareSystems}
 
   \begin{table}
   \caption{Counting Results of Testbed $ 1 $. Lighter colors refer to better performance.}
   \label{tabTestbed1}
   \begin{tabularx}{\columnwidth}{| Y | Y | Y | Y | Y |}
   \hline
   \multirow{2}{*}{\textbf{Real Count}}
    & \multicolumn{2}{c|}{\textbf{\sys{}}}& \multicolumn{2}{c|}{\textbf{Depatal et al.~\cite{Depatla2018}}}\\
    & \textbf{Estimation} & \textbf{Error} & \textbf{Estimation} & \textbf{Error}\\
    \hline\hline
    $ \textbf{0} $ & $ 0 $ & $ 0 $ & $ 0 $ & $ 0 $ \\
    \hline
    $ \textbf{1} $ & $ 1 $ & $ 0 $ & $ 1 $ & $ 0 $ \\
    \hline
    $ \textbf{2} $ & $ 2 $ & $ 0 $ & $ 2 $ & $ 0 $ \\
    \hline
    $ \textbf{3} $ & $ 3 $ & $ 0 $ & $ 2 $ & \cellcolor[gray]{0.8} $ -1 $ \\
    \hline
    $ \textbf{4} $ & $ 4 $ & $ 0 $ & $ 3 $ & \cellcolor[gray]{0.8}$ -1 $ \\
    \hline
    $ \textbf{5} $ & $ 4 $ & \cellcolor[gray]{0.8}$ -1 $ & $ 2 $ & \cellcolor[gray]{0.4} \textcolor{white}{$ -3 $} \\
    \hline
    $ \textbf{6} $ & $ 4 $ & \cellcolor[gray]{0.6} $ -2 $ & $ 3 $ & \cellcolor[gray]{0.4} \textcolor{white}{$ -3 $} \\
    \hline
    $ \textbf{7} $ & $ 8 $ & \cellcolor[gray]{0.8}$ +1 $ & $ 4 $ & \cellcolor[gray]{0.4} \textcolor{white}{$ -3 $} \\
    \hline
   
   \end{tabularx}
  \end{table}

  \begin{table}
   \caption{Counting Results of Testbed $ 2 $. Lighter colors refer to better performance.}
   \label{tabTestbed2}
   \begin{tabularx}{\columnwidth}{| Y | Y | Y | Y | Y |}
   \hline
   \multirow{2}{*}{\textbf{Real Count}}
    & \multicolumn{2}{c|}{\textbf{\sys{}}}& \multicolumn{2}{c|}{\textbf{Depatal et al.~\cite{Depatla2018}}}\\
    & \textbf{Estimation} & \textbf{Error} & \textbf{Estimation} & \textbf{Error}\\
    \hline\hline
    $ \textbf{0} $ & $ 0 $ & $ 0 $ & $ 0 $ & $ 0 $ \\
    \hline
    $ \textbf{1} $ & $ 2 $ & \cellcolor[gray]{0.8} $ +1 $ & $ 3 $ & \cellcolor[gray]{0.6}$ +2 $ \\
    \hline
    $ \textbf{2} $ & $ 4 $ & \cellcolor[gray]{0.6} $ +2 $ & $ 4 $ & \cellcolor[gray]{0.6} $ +2 $ \\
    \hline
    $ \textbf{3} $ & $ 3 $ & $ 0 $ & $ 5 $ & \cellcolor[gray]{0.6} $ +2 $ \\
    \hline
    $ \textbf{4} $ & $ 4 $ & $ 0 $ & $ 5 $ & \cellcolor[gray]{0.8} $ +1 $ \\
    \hline
    $ \textbf{5} $ & $ 6 $ & \cellcolor[gray]{0.8} $ +1 $ & $ 5 $ & $ 0 $ \\
    \hline
    $ \textbf{6} $ & $ 6 $ & $ 0 $ & $ 6 $ & $ 0 $ \\
    \hline
    $ \textbf{7} $ & $ 7 $ & $ 0 $ & $ 6 $ & \cellcolor[gray]{0.8} $ -1 $ \\
    \hline
    $ \textbf{8} $ & $ 9 $ & \cellcolor[gray]{0.8} $ +1 $ & $ 7 $ & \cellcolor[gray]{0.8} $ -1 $ \\
    \hline
    $ \textbf{9} $ & $ 7 $ & \cellcolor[gray]{0.6} $ -2 $ & $ 6 $ & \cellcolor[gray]{0.4} \textcolor{white}{$ -3 $} \\
    \hline
    $ \textbf{10} $ & $ 10 $ & $ 0 $ & $ 5 $ & \cellcolor[gray]{0} \textcolor{white}{$ -5 $} \\
    \hline
   
   \end{tabularx}
  \end{table}
 
  \begin{figure}
   \centering
   \subfloat[Testbed 1\label{figCDF1}]{
     \includegraphics[width=0.9\columnwidth]{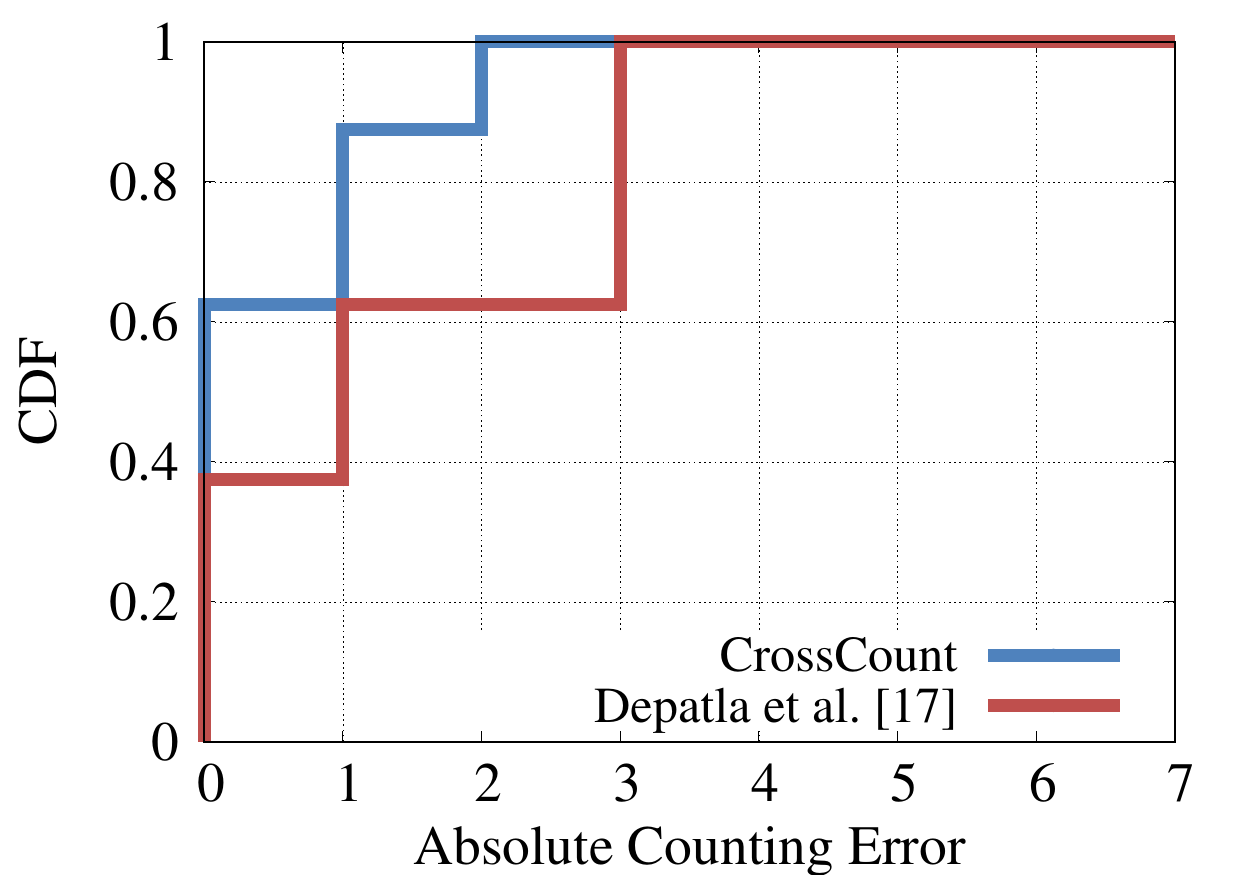}
   }
  
   \subfloat[Testbed 2\label{figCDF2}]{
   \includegraphics[width=0.9\columnwidth]{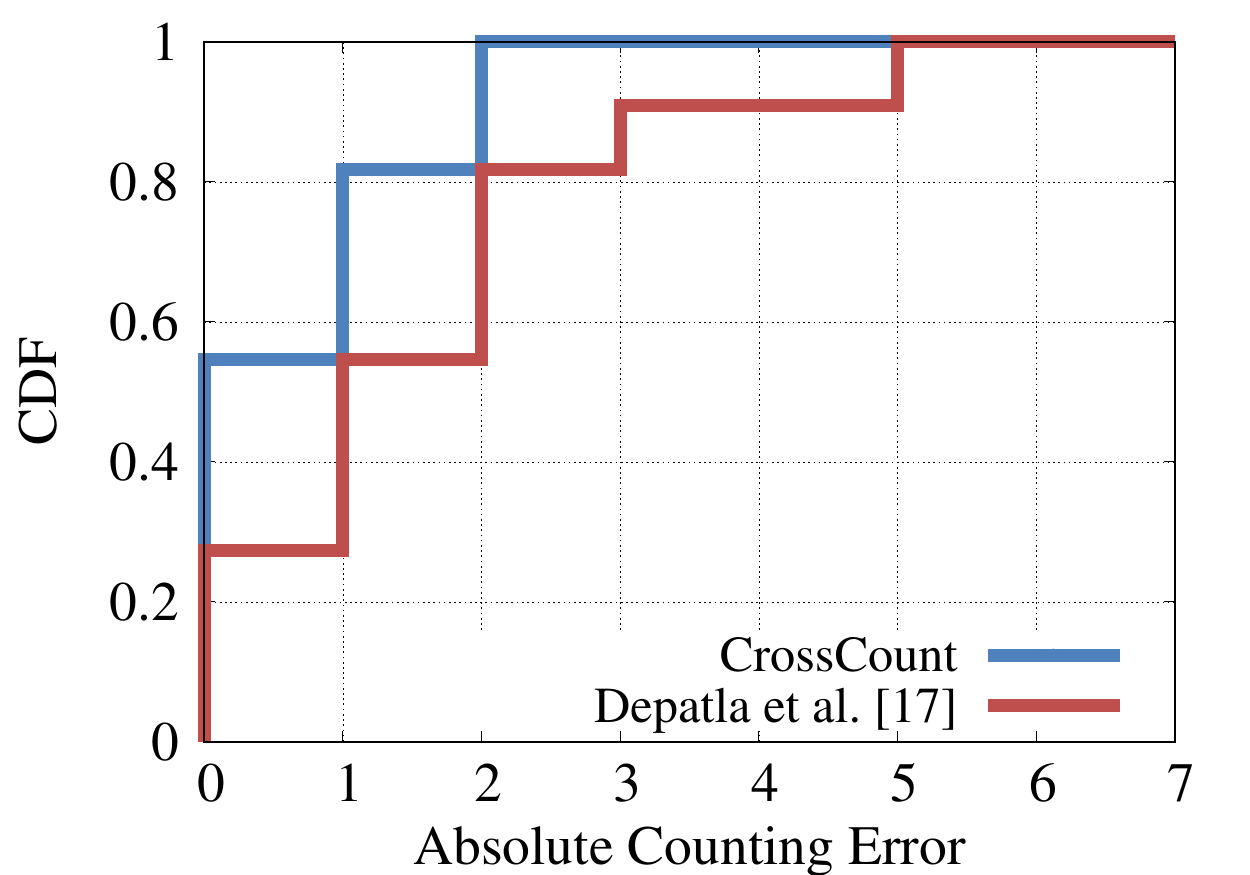}
   }
   \caption{CDF of absolute counting error}
   \label{figCDF}
  \end{figure}
  
  In this section, we compare the \sys{} performance with Depatla et al.~\cite{Depatla2018} as the most recent related work. Moreover, the functionality of the two systems is based on the LoS blockage of a single WiFi link using RSS. For this comparison, $ 10 $ volunteers walked through the testbeds. Due to the limited area of testbed $ 1 $, only $ 7 $ persons participated. The distribution of the absolute counting error $ \varepsilon $ for the two testbeds is reported in Fig.~\ref{figCDF} and detailed in Tables~\ref{tabTestbed1},\ref{tabTestbed2}. In Testbed $ 1 $, \sys{} achieve $ 63\% $ exact count accuracy with only a maximum of $ 2 $-count-difference all the time, while Depatla et al.~\cite{Depatla2018} delivers only $ 39\% $ exact count accuracy. Depatla et al.~\cite{Depatla2018} do not take advantage of the arrival order of blockage events when processing the given counting window unlike \sys{} which consider the whole context information including the arrival order leading to improved performance. The counting estimates are less accurate in Testbed $ 2 $ for both the two systems due to the environment complexity. \sys{} could achieve lower exact accuracy of $ 55\% $ while maintaining a maximum error of $ 2 $ count difference, while Depatla et al.~\cite{Depatla2018} degraded to $ 27\% $ counting accuracy and in some cases it reports $ 5 $ count difference out of $ 10 $ persons. This can be explained by noting that the mathematical model in Depatla et al.~\cite{Depatla2018} is tailored for special cases of testbeds where the WiFi link is in the middle and aligned with the area of interest. These assumptions hold in Testbed $ 1 $, but is not true in Testbed 2, leading to increasing the accuracy improvement for sake of \sys{} over Depatla et al.~\cite{Depatla2018} in the second testbed.
  
  \section{Discussion}
   In this paper, we proposed a novel idea for counting people by classifying their blockage pattern on WiFi links using deep learning. We introduced a data synthesizing  technique that augmented the training set for robust training and provided a lightweight calibration phase. We conduced many experiments that proved the idea and showed that the proposed data generation technique could capture the reality to a reasonable extend. However, the presented system has some limitation that are encouraged to be addressed in the future extensions.
   
   First, \sys{} generated the multiple-person training data by superposing the blockage pattern of a single person. However, further improvement in system performance is expected when simulating the multi-person data as the convolution of RSS signals of single-person. Moreover, the superposition technique could be enhanced by applying left\textbackslash right translation on blockage sequences before being superposed. Second, \sys{} implemented a higher level data noising approach where the blockage events are noised to simulate the RSS changes due to wireless characteristics. However, a lower noising level could be considered by simulating the signal amplitude changes. Finally, \sys{} assumed a casual human motion with non-zero speed while collecting the training and testing sequences. Nonetheless, more sophisticated walking patterns could be investigated, besides, some special scenarios of human movement might be handled, such as when a human stands at the LoS generating a contiguous blocking sequence, among others.

  \section{Conclusion}\label{secConclusion}
   In this paper, we presented the design, implementation, and evaluation of \sys{}: an accurate human counting system based on Recurrent Neural Networks. The system provides different techniques for handling a number of challenges found in the literature such as through-wall signal weakness, labor-intensive data collection, imbalanced training data, high training overhead, high number of data links, and unavailability of CSI data in commodity devices. The main idea is to process the WiFi link blockage inter-arrivals rather than depending on the statistical features extracted from RSS values. By classifying the blockage pattern using an LSTM network, \sys{} achieved superior counting accuracy than the current state-of-art single-link RF-based counting systems.
 
   Currently, we are extending \sys{} in different directions including extending the system to work with multiple links and leveraging CSI information when available.


%



\ifCLASSOPTIONcaptionsoff
  \newpage
\fi


\bibliographystyle{IEEEtran}
\bibliography{references.bib}

%








\end{document}